# Inferring Brain Dynamics via Multimodal Joint Graph Representation EEG-fMRI

*Jalal mirakhorli,*

*jalalmiry@gmail.com*

*ABSTRACT*

*Recent studies have shown that multi-modeling methods can provide new insights into the analysis of brain components that are not possible when each modality is acquired separately. The joint representations of different modalities is a robust model to analyze simultaneously acquired electroencephalography and functional magnetic resonance imaging (EEG-fMRI). Advances in precision instruments have given us the ability to observe the spatiotemporal neural dynamics of the human brain through non-invasive neuroimaging techniques such as EEG & fMRI. Nonlinear fusion methods of streams can extract effective brain components in different dimensions of temporal and spatial. Graph-based analyzes, which have many similarities to brain structure, can overcome the complexities of brain mapping analysis. Throughout, we outline the correlations of several different media in time shifts from one source with graph-based and deep learning methods. Determining overlaps can provide a new perspective for diagnosing functional changes in neuroplasticity studies.*

**Introduction:**

The fMRI and EEG are two of the main non-invasive techniques that allow to measure brain function. An electroencephalogram (EEG) is a test that detects electrical activity in brain using small, metal electrodes attached to scalp. The electrical activity of the brain is caused by the synchronization of a pool of cortical neurons, in particular of pyramidal cells. There is a different electrical charge along the neurons. This difference creates an electric dipole that can be acquired by EEG electrodes, and represented as a series of positive and negative waves.

The fMRI is sensitive to change blood oxygen level dependent (BOLD), that describes the variation in the magnetic status of the red blood cells linked to the hemoglobin oxygenation. There are two forms of hemoglobin, the oxyhemoglobin is the form of hemoglobin loosely combined with oxygen with diamagnetic property whereas the deoxyhemoglobin is the form of hemoglobin that has released its bound oxygen with paramagnetic property. Brain activity in each area causes changes in BLOD in that area of the brain, which is monitored by fMRI techniques as a time series.

The brain is a dynamic graph of interacting neurons that are constantly configured to solve an unexpected hyper-task. These configurations can be supervised or semi-supervised [1]. A multi-modal system can solve complex topologies that are difficult or impossible for an individual stream. Dynamic graph generation aims to generate high-quality structure which are consistent with input time series from both modalities EEG & fMRI. This task requires to simultaneously encode the time series, and effectively leverage the graphs in the decoding process in high order domain. While is hard for typical neural generation models to learn the alignments between source relations.

Numerous studies have shown the efficiency of the simultaneous combination of EEG and fMRI [2]. Directional overlap and convergence in different domains have ensured this combination. Although the EEG and fMRI streams are in different domains and each has its own attention, they have a certain high-order correlation that follows a certain pattern for each task. EEG is high-dimensional in time and fMRI is high-dimensional in space and their fusion has the potential to achieve the potential for high spatiotemporal resolution. These physiological signals are directly captured from the brain cortex and hence it would be advantageous to reflect the mental states of human beings.

EEG and fMRI signals of each brain region are structured time series. Some works have used a stream to represent spatio-temporal graphs [3, 4] that the temporal graph characterizes the dynamics of brain activity at each region and the spatial graph characterizes the functional interaction between different brain regions. Which usually analyzes spatio-temporal in rotation or uses partial graph correlation [5, 6], while either lose good temporal information or ignore some spatial data. To address these issues, we suggest cross model to incorporate spatio-temporal convolution on functional & temporal connectivity graphs. Therefore, simultaneous acquisition of EEG and fMRI data covers each other's weaknesses, while improving system noise response especially similar artifacts. EEG has reliable temporal accuracy and spectral resolution of the brain electrophysiology, while fMRI is suitable for spatial analysis of blood flow changes [7, 8]. Furthermore, by simultaneously obtaining data, the function of the brain can be examined in the same states, with different media in the same experimental conditions. Including things like the level of consciousness and concentration that are dynamically changing in humans [9, 10].

There are generally two methods for fusion that are usually biased according to physiological assumptions. A) Asymmetrical approaches, the data extracted from one stream is used to limit or predict another stream. And according to the priority of using stream information, the analysis form is divided into two types. One fMRI-informed EEG, in which fMRI spatial information is used for the source reconstruction of the EEG signals [11] and in the other type, EEG-informed fMRI, the temporal information obtained from EEG is used to localize and infer the fMRI correlations to specific neuronal events [12, 13]. In these approaches, the second modality is biased based on the first modality. And small changes in the first stream can eliminate the effective feature in the second stream and only some features are considered. B) Symmetrical approaches, where a wide range of features of the two modalities are jointly examined, and in other words, the features have the same weight in a similar analysis [14]. This approach is classified into two groups: data & model- driven methods [15, 16, 17, and 18].

In the data-driven method, a blind search is performed to detect common patterns between streamers by linear and nonlinear methods without considering the physiological relationships of EEG and fMRI [10]. On the other hand, the model-driven method is based on hypothetical neurophysiological that describe the connection between the neuronal generators and the bioelectrical hemodynamic responses so there is a possibility of losing or not considering effective connectivity [19]. As a result, unlike the first method, there is a possibility of losing or not considering effective communication. Therefore, data-driven model is preferred in situations where accurate biophysical modeling and prediction of brain events are not possible [20].

Given the complex data structure in different imaging modalities as well as networked organizational structure of the human brain, novel learning methods based on graphs inferred from imaging data, graph regularizations for the data, and graph embedding of the recorded data, have shown great promise in modeling the interactions of multiple brain regions, information fusion among networks derived from different brain imaging modalities, latent space modeling of the high dimensional brain networks, and quantifying topological novel neuro-biomarker in temporal dependencies and attention structures [21]. In the meantime, attention mechanisms can be used to optimize the system. Attention mechanisms are divided into two main types: soft-attention and self-attention mechanisms. Soft-attention mechanisms allow the model to learn the most relevant parts of the input sequence during training. Soft-attention mechanisms are end-to-end approaches that can be learned by gradient-based methods. Self-attention mechanisms incorporate the attention mechanism into the propagation steps by modifying the convolution operation [22, 23, 24, and 25].

In this study use a graph topology [26] such as the brain topology and analyze it in terms of time and patterning of brain functions to model and configure a multi-modalities system. Using the graph model, it is possible to find dependencies between brain areas (adjust correlation) more easily, and find possible solutions using sub-graphs, just like the brain. Graph analysis focuses on tasks such as node classification, link prediction, and clustering. Graph Convolutional Neural Networks (GCNNs) are deep learning based methods that operate on graph domain in modeling unstructured and structured relational data including brain signals (fMRI and EEG) to investigate structure-function relationships in the human brain [21, 27].

Components and correlated areas of the whole-brain are obtained by analyzing cortical signals, possibly on subcortical, into a graph network of coupled regions of interest (ROIs), or by approximating the cortical signals as a bi-dimensional manifold. The model that will be described is capable of learning the sparse representation and extracts graph features jointly, providing the flexibility to choose between individual and group level explanations. Here, instead of using a prior graph the algorithm focuses on the influences of relationships to introduce inter-intra GCNN model and jointly learn the structure and connection weights to optimize task-related learning of fMRI and EEG while a self-attention mechanism updates the impact of node connections. However, combining modalities may uncover hidden relationships. Thus, proposed here a graph joint representation learning framework to deal with the challenge. Finally, the degree of correlation between the two streams is determined.

In this article introduces a new attention-based multi-scale GCNN model, aiming to feature extraction and supervised classification. GCNN is a unified, which simultaneously learns the discriminative fMRI and EEG feature representations, classifies the driving brain states, and exploits the functional topological relationship of electrodes and ROI. The model starts with temporal convolutions to learn frequency filters from the preprocess fMRI and EEG data, in which the kernel size is simply determined by the sampling rate of data, and hence, the tuning of kernel size is not necessary that feed attention-based multi-scale method to reveal the correlation between brain connections.

**Method:**

**Graph analysis:** A standard convolution for regular grids is clearly not applicable to general graphs. There are two basic approaches currently exploring how to generalize GCNNs to structured data forms. One is to expand the spatial definition of a convolution. Spatial GCNs imitate the Euclidean convolution on grid data to aggregate spatial features between neighboring nodes [28]. In fact, spatial based GCNs define a graph convolution operation based on the spatial relationships that exist among the graph nodes. And the other is to manipulate in the spectral domain with graph Fourier transforms. These methods can be decomposed into spectral bases associated with graph-level information and include graph filtering-based approaches according to spectral graph theory. [29, 30]. Spectral method has proven to be very effective considering the purposes of this article including extracting effective links [31, 32, and 33]. So the convolution operator should behave like a low pass filter by smoothing the feature of each node on the graph using node features in its neighborhood.

An undirected and connected graph can be defined as G = {V, E, W}, in which V represents the set of nodes, parcellation of cerebral cortex or electrode, with the number of |V| = N and E denotes the set of edges connecting these nodes. Let $W \in R^{N \times N}$ denote an adjacency matrix describing the connections between any two nodes in V, in which the entry of W in the i-th row and j-th column, denoted by $w_{ij}$, measures the importance of the connection between the i-th node and the j-th one while assigning a brain signal $x \in \mathbb{R}^{N \times T}$, i.e. a short time series with duration of $T$, to each of $N$ brain regions or electrode. methods k-nearest-neighbor (k-NN) and the distance function method are used to build graphs and determine the entries $w_{ij}$ of the adjacency matrix with only connecting each brain region to its 8 neighbors with highest connectivity whereas a distance function would be the Gaussian kernel function [33, 34].

The spectral analysis of the graph signals relies on the graph Laplacian, which maps the signal distributions from the spatial domain to the graph spectral domain and decomposes the signals into a series of graph modes with different frequencies, also called graph convolution. Specifically, the normalized graph Laplacian matrix is defined as:

$$L = D - W \in R^{N \times N}, \quad (1)$$

Where L denote the Laplacian matrix of the graph G, $D \in R^{N \times N}$ is a diagonal matrix and the i-th diagonal element can be calculated by $D_{ii} = \sum_j w_{ij}$.

The convolution operator is shown for a given spatial signal $x \in R^N$ as follows:

$$\hat{X} = U^T X, \quad (2)$$

Where $\hat{X}$ denotes the transformed signal in the frequency domain, U is an orthonormal matrix that can be obtained via the singular value decomposition (SVD) of the graph Laplacian matrix $L = U\Lambda U^T$, in which the columns of $U = [u_0, \cdots, u_{N-1}] \in R^{N \times N}$ constitute the Fourier basis, and $\Lambda = \text{diag}([\lambda_0, \cdots, \lambda_{N-1}])$ is a diagonal matrix. And the inverse conversion to this form is shown:

$$x = UU^T x = U\hat{X}, \quad (3)$$

As a result, the convolution of two signals x and y on the graph $*_g$ is as follows:

$$x *_g y = U((U^T x) \odot (U^T y)), \quad (4)$$

Where $\odot$ denotes the element-wise Hadamard product. $g(\cdot)$ can be a filtering function such that a signal x filtered by $g(L)$ can be expressed as:

$$y = g(L) x = g(U\Lambda U^T) x = U g(\Lambda) U^T x, \quad (5)$$

Where $g(\Lambda) = \text{diag}(g(\lambda_0), \ldots, g(\lambda_{N-1}))$ denotes a diagonal matrix with N spectral filter coefficients. In fact, the convolution between the graph signal $x \in \mathbb{R}^{N \times T}$ and the vector of $U g(\Lambda)$ based on graph G is defined as their element-wise Hadamard product in the spectral domain:

$$y = g(L) x = U (g(\Lambda)) \odot (U^T x) = U (U^T (U g(\Lambda))) \odot (U^T x) = x *_g (U g(\Lambda)). \quad (6)$$

K is the kernel size of graph convolution, which determines the maximum radius of the convolution from central nodes. Since it is difficult to directly calculate the expression of $g(\Lambda)$, here we only consider the Chebyshev polynomials of the diagonal matrix of Laplacian eigenvalues. In order to localize the filters in space and reduce their computational complexity, the truncated expansion of the K-order Chebyshev polynomial is used to approximate the filter [35]. A polynomial of order K yields strictly K-localized filters and as mentioned before, a convolution in the graph spatial domain corresponds to a multiplication in the graph spectral domain.

First, Under the K order Chebyshev polynomials framework, $g(\Lambda)$ is approximated by:

$$g(\Lambda) = \sum_{k=0}^{k-1} \theta_k T_k(\Lambda). \quad (7)$$

Where $\theta_k$ is the coefficient of Chebyshev polynomials, and $T_k(x)$ can be recursively calculated according to the following recursive expressions:

$$T_0(x) = 1, T_1(x) = x, \quad (8)$$

$$T_k(x) = 2xT_{k-1}(x) - T_{k-2}(x), \quad k \geq 2. \quad (9)$$

If $\lambda_{max}$ denote the largest element among the diagonal entries of $\Lambda$ and denote the normalized $\Lambda$ by $\Lambda_n = 2\Lambda/\lambda_{max} - I_N$, such that the diagonal elements of $\Lambda_n$ lie in the interval of $[-1, 1]$, where $I_N$ is the $N \times N$ identity matrix. According to the above equations, the graph convolution operation is obtained as follows:

$$y = U\ g(\Lambda)U^T x = \sum_{k=0}^{k-1} \theta_k T_k(\Lambda_n)x. \qquad (10)$$

In fact, graph convolution operation represents combination of the convolutional results of x with each of the Chebyshev polynomial components or Filtering of a signal x with a K-localized filter can be shown as Equation 10.

In the first phase, the effective connections of each modality (fMRI and EEG) must be identified. For this purpose, the architecture of convolutional layers is based on the Class Activation Maps [36]. This architecture highlight the most important connections in the model's classification process. Therefore, the most important or salient nodes for the classification correlate well with characteristic graph features of the corresponding functional type. It has been shown that if a global average pooling (GAP) layer is used at the end of the neural network instead of a fully-connected layer, will result in an excellent node localization. Even though this method in this case was trained for classification, by looking at the areas where the network paid attention, it achieved decent results in node localization. To do this, the network should be trained with a GAP layer. After that layer, there are a fully-connected network followed by a softmax layer, while conditioned on a specific class. In CAM generates saliency map by taking the weighted average of layer output channels using the weights of the fully-connected network that's connected to the output class. The weights of the dense layer can linearly map onto the corresponding feature maps to generate a class activation map showing the salient nodes in the graph. To obtain salient connections and nodes among all subjects, an argmax operator is used that returns the effective connections and nodes in each class, and this salient map is used to fuse the modalities in the next step. Details are shown in Figure 1.

In the following, the correlations that can be obtained from the content of EEG electrodes based on inter-node interactions and ROI from fMRI are examined.

Given a set of EEG and fMRI features vector of n nodes $v_f = \{\vec{f_i}\}_{i=1 \text{ to } n}$ of n nodes, which has already been obtained in GCNN, the corresponding node correlations are computed as:

$$C^{ij} = \text{sim}(\vec{f_i}, \vec{f_j}). \qquad (11)$$

Where sim(·) denotes the cosine similarity [37].

The result of the GCNN in the first phase is multiple saliency maps, nodes and related features of affective fMRI and EEG. Here, the number of effective nodes extracted from saliency map in the fMRI modality is considered as a reference. The goal of joint representation learning is to learn a distributed and low dimensional embedding of each ROI, which is denoted as $J$,

$$\mathcal{E} = \{\vec{e_0}, \vec{e_1}, \ldots, \vec{e_{|n|}}\}, \vec{e} \in \mathbb{R}^d, \forall\ e \in \mathcal{E}, \qquad (12)$$

Where $\vec{e_i}$ is the embedding of *i-th* ROI, and *d* is the embedding size. In the d-dimension embedding space, the ROI correlations revealed by both the fMRI and EEG features in the saliency maps are preserved. First, we introduce the modeling of multi-view correlations between ROI from multiple saliency maps. ROIs are related to each other from multiple aspects. For example, to make decisions, networks of neurons are involved at different points, some as sensors and some as decision makers, or parts of the brain that form a network together by default. Alternatively, according to the inherent attributes of regions such as default mode or wave propagation between electrodes in EEG mode, adjacent regions show high correlations due to similar functionalities. To learn robust and comprehensive representations of ROIs, we have to consider node correlations from multiple views jointly.in here, we construct five types of node correlations based on EEG components and ROIs in fMRI.

In the following, the correlations that can be obtained from the content of EEG electrodes based on inter-node interactions and ROI from fMRI are examined. For this purpose, first learns node representations using node correlations from single view and graph attention network (GAT) [38] is employed to graphs from saliency maps to learn representations of vertices. GAT applies attention mechanism on graph-structured data. It updates the representation of a vertex by propagating information to its neighbors, where the weights of its neighbor vertices is learned by attention mechanism automatically. Given the input vertex feature *M*, a GAT layer updates the vertex representations by following steps:

$$M = \{\vec{f_0}, \vec{f_1}, \ldots, \vec{f_{|n|}}\}, \vec{f} \in R^F, \forall \vec{f} \in M, \qquad (13)$$

$$e_{ij} = \exp(\text{ReLU}(\vec{a}^T, [W\vec{f_i} \| W\vec{f_j}])). \qquad (14)$$

Where (j, i) denotes trace an edge from a node j to a node i and n is the number of nodes. $e_{ij}$ indicate the importance of node j's features to node i and W and $\vec{a}$ are learnable parameters, ‖ is the concatenation operation. To make coefficients easily comparable across different nodes, in here normalize them across all choices of j using the softmax function:

$$\alpha_{ij} = \text{softmax}_j (e_{ij}) = \frac{\exp(e_{ij})}{\sum_{k \in N_i} \exp(e_{ik})}, \qquad (15)$$

$$\vec{f_i} = \sigma \left( \sum_{j \in N_i} \alpha_{ij} W\vec{f_j} \right), \qquad (16)$$

Apply multi-head attention mechanism in each GAT layer as suggested [38] to enhance the performance. Two consecutive layers of the GAT are used to apply to the five graphs obtained from the saliency map and at its output show vertexes and nodes representations as $\mathcal{E}_\beta, \mathcal{E}_\delta, \mathcal{E}_\gamma, \mathcal{E}_\theta, \mathcal{E}_\alpha, \mathcal{E}_{fMRI}$.

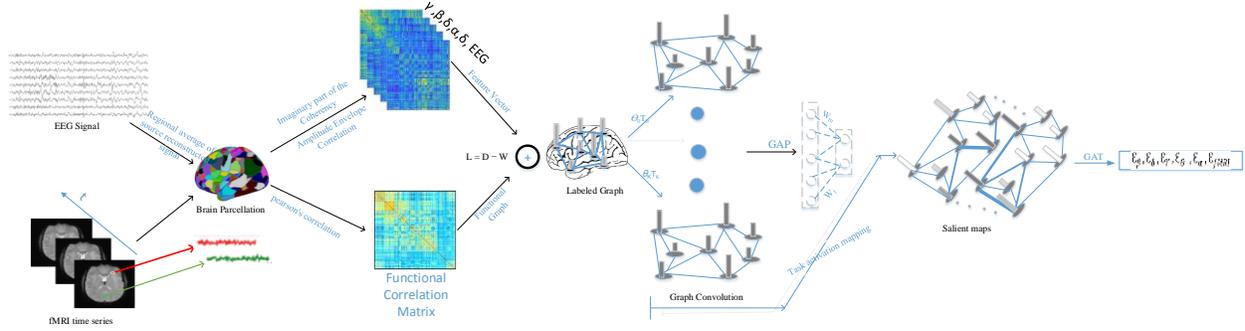

Fig. 1: The overview of the pipeline. fMRI & EEG signals are parcellated by an atlas and transferred to graphs. Then, the graphs are sent to GNN, which gives the prediction of specific tasks, and the salient maps are obtained for each modality. The salient brain regions feed GAT for find vertexes and nodes that are informative to the prediction-related communities.

**Cross- Attention Adaptation:**

Although the features are obtained from different molarities, but they have the same origin then it is reasonable to consider the feature dependencies or correlations for more accurate relationship modeling between nodes of graphs. Also, according to different semantic features, representation of a central entity is generated in semantic representation level. The first part aim to minimize the distance between the features of different modalities in order to mitigate the performance drop caused by the domain gap. Due to such a high correlation of features from multi-view and improving the learning process on each perspective, in here apply multi-head attention [39]. This method is leveraged to stabilize the training process and aggregate semantic features of a central entity in different view relation features. The first step in calculating self-attention is to create two matrixes from each of the view's input matrixes. So for each view, create a Query matrix,Q, and a Key matrix ,K. These matrixes are created by multiplying the embedding by two matrices that trained during the training process. Given the representations from M different modalities as $\{\mathcal{E}_i \in R^{n \times d}\}$, i =1 to M. So for each presentation, $\mathcal{E}_i$, can define a key matrix $K_i \in R^{n \times k}$, a query matrix $Q_i \in R^{n \times k}$, a value matrix $V_i \in R^{n \times k}$ and trainable weight matrices $W_k$, $W_q$, $W_v$ respectively, with it as follows:

$$K_i = \mathcal{E}_i W_k, Q_i = \mathcal{E}_i W_q, V_i = \mathcal{E}_i W_v. \qquad (17)$$

Then, can define content-based self-attention and propagate information among all views as follows: as:

$$[A_i]_{i=1 \text{ to } M} = \text{softmax}\left(\left[\frac{Q_i K_i^T}{\sqrt{k}}\right]\right) V_i, \quad i = 1 \text{ to } M \qquad (18)$$

Since $[A]_i = [A_1, A_2,\ldots, A_M] \in R^{i \times M}$ is a series of output vectors of Q, K , and V , therefore, A is a series of output vectors of $\mathcal{E}_i$, and derive the following equation:

$$\dot{\varepsilon}_i = \sum_{i=1}^{M} A_i \varepsilon_i. \tag{19}$$

Where, $\dot{\varepsilon}_i$ is the new i-th view representation generated by the self-attention module. $\dot{\varepsilon}_i$ is the output of self-attention layer and to use it in the learning process and the percentage of participation i-th view representation, define $\breve{\varepsilon}_i$ as follows:

$$\breve{\varepsilon}_i = \rho \dot{\varepsilon}_i + (1-\rho)\varepsilon_i, \quad 0 < \rho < 1. \tag{20}$$

Multi-view Fusion

Multi-view fusion is applied to saliency map as treated with considering the feature quality of each view and finally the preservation of correlated features. However, the strategy is problematic in some cases where the saliency map of some views may have artifact or noise. To solve this problem, in this here present a weight learning network as fusion layer to learn an adaptive weight for different views as follows:

$$\varepsilon_f = \omega_i \varepsilon_i + \sum_{i=1}^{M} \omega_i \varepsilon_i. \tag{21}$$

Where $\omega_i$ is the weight of i-th view, which is learned by a single layer MLP network with the i-th embedding as input and regresses M weights $\omega_i$.

In order to enable the learning of the multi-view fusion layer, $\varepsilon_f$ will be a candidate in the learning objective of each view and update the representation of each view as:

$$\breve{\varepsilon} = \frac{\varepsilon_f + \dot{\varepsilon}_i}{2}. \tag{22}$$

By using the representation of each view, $\breve{\varepsilon}$, into a joint learning module, obtain node embedding, $\breve{\varepsilon}_\beta, \breve{\varepsilon}_\gamma, \breve{\varepsilon}_\delta, \breve{\varepsilon}_\alpha, \breve{\varepsilon}_\Theta, \breve{\varepsilon}_{fMRI}$, on which design learning tasks. Details are shown in Figure 2.

**Learning Network:**

To effectively training our model, in this case propose a simple network consisting of one tasks based on node relation reconstruction that is the graph geometry embedding while considers the cross-view node consistency. The proposed weight learning network can be joined with the node estimation network for end-to-end training without enforcing supervision on the weights. In this process, the network learns node embedding to reserve the node similarity in terms of different node attributes, in this task reconstructs node correlations based on corresponding embedding. Take $EEG_\beta$ attributes as example, the learning objective is defined based on $c_\beta^{ij}$ and $\breve{\varepsilon}_\beta = \{e_\beta^i\}_{i=1}$ to n as follow:

$$\zeta_\beta = \sum_{ij}(c_\beta^{ij} - e_\beta^{iT} e_\beta^{j})^2, \tag{23}$$

In this way, the final learning objective is:

$$\zeta_{total} = \zeta_\beta + \zeta_\alpha + \zeta_\Theta + \zeta_\gamma + \zeta_\delta + \zeta_{fMRI}. \tag{24}$$

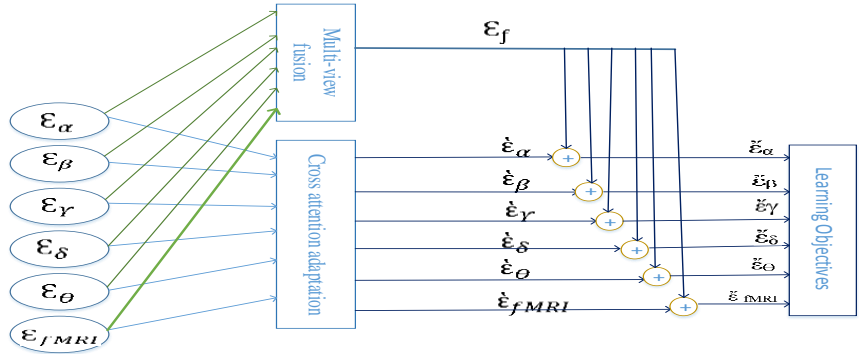

Fig 2. Multi-view and Cross attention adaptation processing architecture.